\documentclass{article}
\usepackage{hiph-preprint}
\newcommand{\RCP}{R_{\rm CP}}
\newcommand{\pt}{p_{\tt T}}

\newcommand{\av}[1]{\left\langle #1 \right\rangle}
\newcommand{\gev}{\mathrm{GeV}}
\newcommand{\sqrtsNN}{\sqrt{s_{\scriptscriptstyle{{\rm NN}}}}}
\newcommand{\Kzs}{{\rm K^0_S}}
\newcommand{\La}{\Lambda}
\newcommand{\Al}{\overline\Lambda}
\usepackage{pennames}
\usepackage{graphicx}
\usepackage{amssymb}
\volnumber{22} \issuenumber{1} \edyear{2005}                             
\frompage{000} \topage{000}                                              
\recrevdate{\today}                                                      
\def\rightmark{$\RCP$\ of strange particles at the SPS}
\def\leftmark{G.E. Bruno and A. Dainese}
 
\title{First measurement of the strange particles  
       $\RCP$ nuclear modification factors in heavy-ion collisions  
       at the SPS} 
\authors{ 
{G. E. Bruno$^1$ and A. Dainese$^{2}$ for the 
NA57 Collaboration\footnote{For the full author list
see Appendix ``Collaborations'' in this volume.} %
\index{Bruno, G.E.} 
\index{Dainese, A.} 
}\\[2.812mm]
{\normalsize
\hspace*{-8pt}$^1$ Dipartimento IA di Fisica dell'Universit\`a e 
                   del Politecnico di Bari and INFN,\\ 
Bari, Italy\\[0.2ex] 
\hspace*{-8pt}$^2$ Dipartimento di Fisica dell'Universit\`a 
degli Studi di Padova and INFN,\\ 
Padova, Italy
}}
 
\abstract{The NA57 experiment 
has measured the $\pt$\ distributions of \PKzS, \PgL, and
\PagL\ particles 
in fixed-target Pb--Pb interactions  
at $\sqrtsNN=17.3~\gev$, as a function of the collision centrality. 
In this paper we study the central-to-peripheral nuclear modification 
factors and compare them to other measurements and to theoretical predictions.}

\keyword{nucleus--nucleus collisions; strange particles}

\PACS{12.38.Mh; 25.75.Nq; 25.75.Dw}
 
\makeindex
\begin{document}
 
\maketitle

\section{Introduction}

At the Relativistic Heavy Ion Collider (RHIC), 
the central-to-peripheral nuclear modification factor
\begin{equation}
\RCP(\pt) = {\av{N_{\rm coll}}_{\rm P} \over \av{N_{\rm coll}}_{\rm C}}\times
\frac{{\rm d}^2 N_{\rm AA}^{\rm C}/{\rm d}\pt{\rm d} y}{{\rm d}^2 N_{\rm AA}^{\rm P}/{\rm d}\pt{\rm d} y}
\label{eq:rcp}
\end{equation}
has proven to be a powerful tool for the study of parton propagation 
in the dense QCD medium expected to be formed in nucleus--nucleus collisions
 (see, e.g.,~\cite{starRcpk0la}).
At high $p_T$ ($> 7~\gev/c$), 
$\RCP$ is found to be suppressed by a factor 3--4 with respect 
to unity (at $\sqrtsNN=200~\gev$), for all particle species;
this is interpreted as a consequence of parton energy loss in the medium, 
prior to fragmentation outside the medium~\cite{white}. 
Energy loss would predominantly 
occur due to additional gluon radiation, with a 
probability that depends on the density of colour charges in the 
medium~\cite{gyulassywang,bdmps}. 
High-$\pt$ partons are then believed to fragment 
in a similar fashion as in pp collisions 
($\pt^{\rm hadron}=z\,\pt^{\rm parton}$; $z<1$).
At intermediate $\pt$ ($2$--$5~\gev/c$), instead, 
after energy loss, partons would  
hadronize `inside' the dense partonic system
via the mechanism of parton coalescence~\cite{coalth}: 
a quark-antiquark pair or 
three (anti)quarks, close in phase-space, would coalesce into 
a meson or baryon, 
with $\pt^{\rm meson(baryon)}\approx 2(3)\,\pt^{\rm quark}$. 
This is predicted to originate 
a pattern of larger $\RCP$ for 
baryons relative to mesons. The pattern has been observed, for instance,
 by the STAR
Collaboration for $\Lambda$, $\Xi$, and $\Omega$  relative to 
$\rm K^0_S$, $\rm K^{0\star}$, and $\phi$~\cite{starRcpk0la,starRcpXiOm,starRcpphi}.  

The study of $\RCP$ and of its particle-species dependence,
in particular the meson/baryon dependence, 
at top SPS energy allows to test for these 
phenomena at an energy smaller by about one order of magnitude  
($\sqrtsNN=17.3~\gev$). While measurements of the $\pi^0$ $\RCP$
by the WA98 Collaboration, supporting the presence 
of parton energy loss effects, were
 published already in 2002~\cite{wa98}, 
the first results on the 
particle-species dependence (unidentified 
negatively charged hadrons, \PKzS, \PgL, and
\PagL) have been  
reported by the NA57 Collaboration in~\cite{rcppaper}.  

\section{Data analysis}
Strange particles are reconstructed using their decay channels
into charged particles, measured in the NA57 silicon pixel telescope: 
$\Kzs \to \pi^+\pi^-$, $\La\to\pi^-{\rm p}$, and
$\Al\to\pi^+\overline{\rm p}$.
The selection procedure is described in detail
in~\cite{rcppaper}. The main criteria are the following:
(a) the two decay tracks are compatible with the hypothesis of having a common
    origin point;
(b) the reconstructed secondary vertex is well separated from the
      target;
(c) the reconstructed candidate points back to the primary vertex position.
The acceptance covers about one
unit of rapidity around mid-rapidity.

The collision centrality is determined from the charged particle
multiplicity $N_{\rm ch}$,
measured by a dedicated silicon detector~\cite{wounded,mult2004}.
The sample of collected events is subdivided in centrality classes, 
with $N_{\rm ch}$ limits corresponding to given
fractions of the \mbox{Pb--Pb} inelastic cross section.
The covered centrality range is 0--53\% $\sigma^{\rm Pb-Pb}$. 
For each class the average
number of participants, $\av{N_{\rm part}}$, and of binary 
collisions, $\av{N_{\rm coll}}$, are calculated
from the Glauber model, after a wounded-nucleon model 
fit ($N_{\rm ch}\propto N_{\rm part}$) to the $N_{\rm ch}$
distribution~\cite{wounded}. 

\section{Results and comparisons}

\begin{figure}[!t]
\begin{center}
\includegraphics[width=\textwidth]{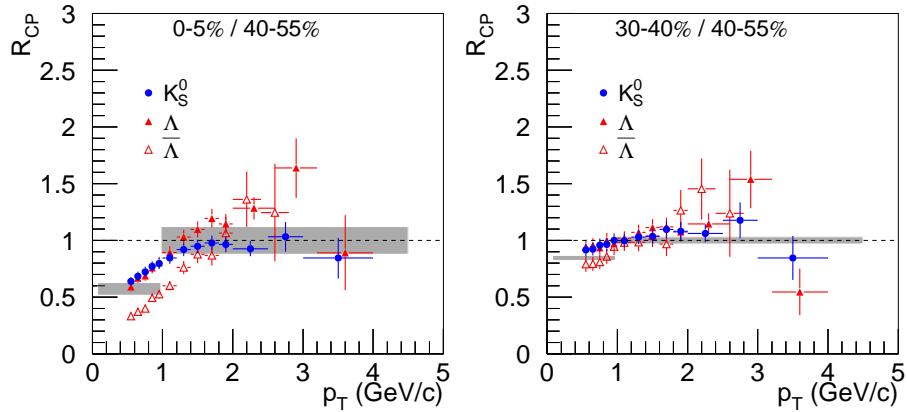}
\caption{$\RCP$ ratios 
          for singly-strange particles in 
           Pb--Pb collisions at $\sqrtsNN=17.3~\gev$~\cite{rcppaper}.
           The error bars show the quadratic sum of the statistical errors
            and the $p_T$-dependent systematic errors ($<7\%$). 
            Shaded bands centered
            at $\RCP=1$ represent the systematic error due to the uncertainty
            in the ratio of the
            values of $\av{N_{\rm coll}}$ in each class; shaded
            bands at low $\pt$ represent the values expected for
            scaling with the number of participants, together with their
            systematic error. }
\label{fig:Rcp}
\end{center}
\end{figure}

Figure~\ref{fig:Rcp} shows the results for 0--5\%/40--55\% and 
30--40\%/40--55\% $\RCP$ nuclear modification factors.
At low-$\pt$, $\RCP$ scales with the number of participants for 
all particles except the $\overline\Lambda$. With increasing $\pt$,
$\rm K_S^0$ mesons reach values of $\RCP\approx 1$: for the most central 
class we do not observe the enhancement above unity that was measured in 
proton--nucleus relative to pp collisions
(Cronin effect~\cite{cronin}). An enhancement is, 
instead, observed for strange baryons, $\Lambda$ and $\overline\Lambda$,
that reach $\RCP\simeq 1.5$ at $\pt\simeq 3~\gev/c$.

\begin{figure}[!t]
\begin{center}
   \includegraphics[width=.48\textwidth]{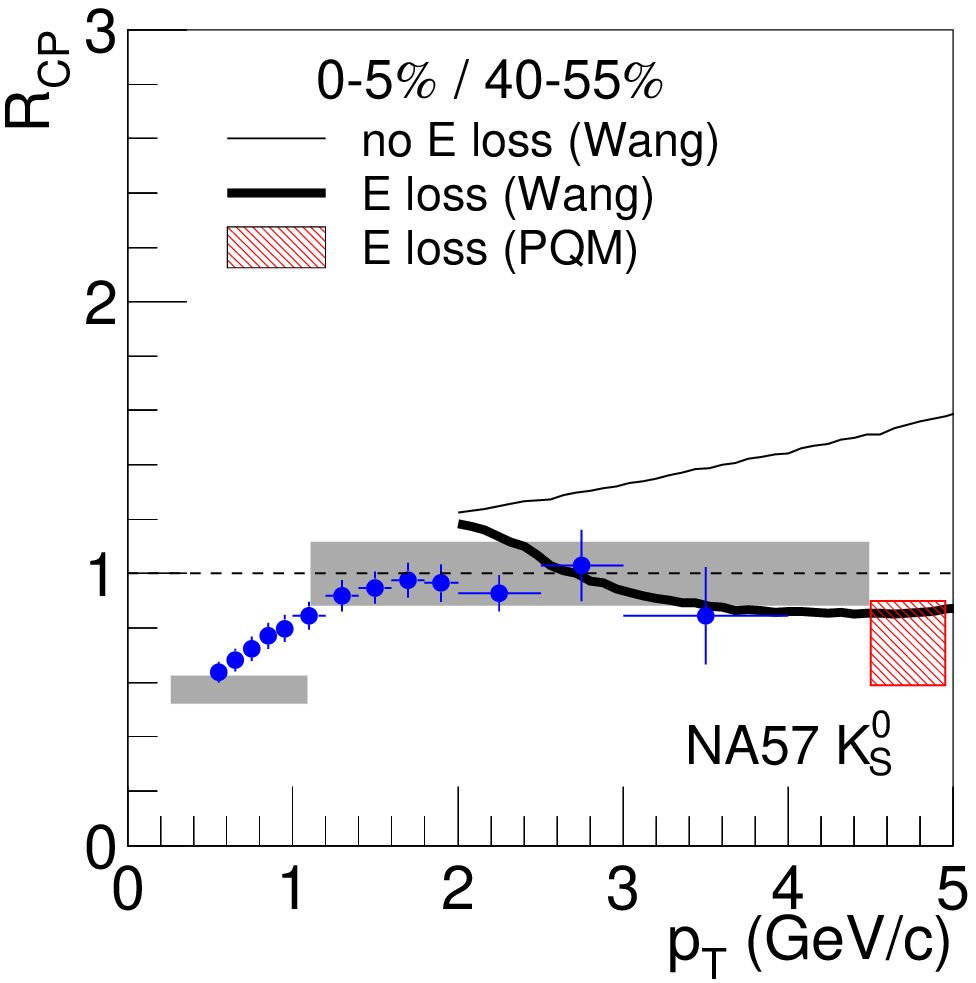}
   \includegraphics[width=.48\textwidth]{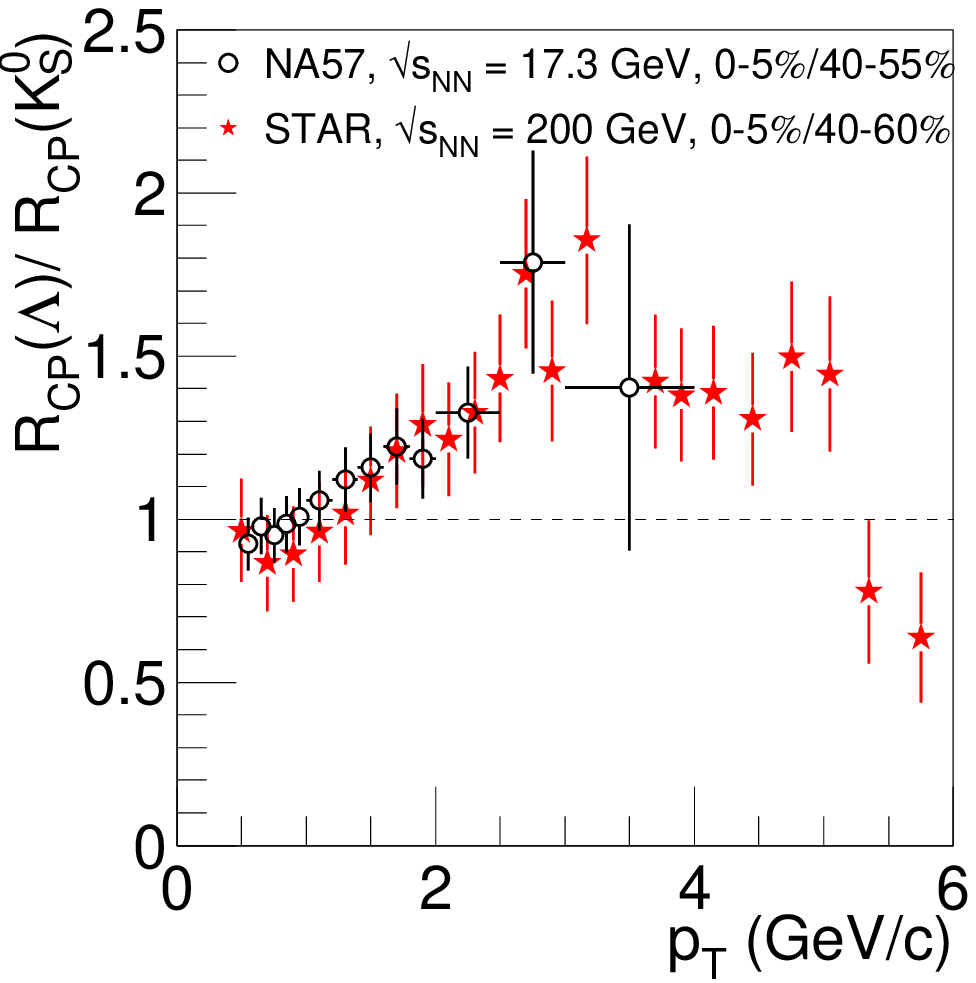}
   \caption{Left: $\rm K^0_S$ $\RCP(\pt)$\ 
            compared to predictions~\cite{wang,pqm}
            with and without energy loss. 
            Right: ratio of $\Lambda$ $\RCP$ to $\rm K^0_S$ $\RCP$,
            as a function of $\pt$, at the  
            SPS (NA57 at $\sqrtsNN=17.3~\gev$) and at 
            RHIC (STAR at $\sqrtsNN=200~\gev$~\cite{starRcpk0la,lamont}).}  
   \label{fig:RcpTh}
\end{center}
\end{figure}

In Fig.~\ref{fig:RcpTh} (left)
we compare our $\rm K^0_S$ data to predictions (X.N.~Wang)
obtained from a perturbative-QCD-based
calculation~\cite{wang}, including (thick line) or
excluding (thin line) in-medium parton energy loss. The initial
gluon rapidity density of the medium
was scaled down, from that
needed to describe RHIC data, according to the decrease by
about a factor 2 in the charged multiplicity.
The data are better described by the curve that does include energy loss. 
Also the prediction of a
second model of parton energy loss (PQM),
that describes several energy-loss-related observables at RHIC
energies~\cite{pqm},
is in agreement with the value reached at high $\pt$ by our ${\rm K^0_S}$ data.

Figure~\ref{fig:RcpTh} (right) shows the ratio 
of $\Lambda$ $\RCP$ to ${\rm K^0_S}$ $\RCP$, as measured from our data
and by STAR
at $\sqrtsNN=200~\gev$~\cite{starRcpk0la,lamont}
(note that $\Lambda+\overline\Lambda$ are considered by STAR and that the 
centrality range used for the peripheral class is slightly different 
in the two experiments).
The similarity of the $\Lambda$--$\rm K$ pattern
to that observed at RHIC 
may be taken as an indication for coalescence  
effects at SPS energy. Note also that a qualitatively similar baryon--meson
relative pattern at the SPS was presented by NA49 for protons and 
charged pions~\cite{na49}.

\section{Conclusions}
\label{concl}
We have presented 
the first results on central-to-peripheral nuclear modification factors 
for strange particles at the SPS energy $\sqrtsNN=17.3~\gev$.
The comparisons of our data to theoretical calculations and STAR data 
at $\sqrtsNN=200~\gev$
 suggest a scenario with moderate in-medium partonic 
energy losses and a recombination-induced baryon/meson effect, 
similar to that observed at RHIC energy.

\vfill\eject
\end{document}